\documentstyle[prl,aps]{revtex}
\let\section=\subsection  \let\subsection=\subsubsection

\def\be{\begin{equation}}
\def\ee{\end{equation}}
\def\bea{\begin{eqnarray}}
\def\eea{\end{eqnarray}}

\input{epsf}

\begin{document}

\begin{center}
{\large {\bf A Model to Describe Transport Properties in $%
Bi_2Sr_2(Ca_zPr_{1-z})Cu_2O_{8+y}$ }}\\[8mm]{E.C. Bastone \footnote{%
work partially developed at DCNat - Departamento de Ci\^{e}ncias Naturais,
FUNREI - Funda\c{c}\~{a}o de Ensino Superior de S\~{a}o Jo\~{a}o del Rei. Pra%
\c{c}a Dom Helvecio, CEP:36.300-000,S\~{a}o Jo\~{a}o del Rei - M.G. Brazil},
A.S.T. Pires and P.R. Silva\\[5mm]{\small {\it Departamento de F\'{\i}sica,
Instituto de Ci\^{e}ncias Exatas,\\Universidade Federal de Minas Gerais,\\%
Belo Horizonte, CEP 30.123-970, C.P. 702, MG, Brazil \\e-mail:
erika@fisica.ufmg.br\\[8mm]}}}
\end{center}

\centerline{{\bf ABSTRACT}}
\begin{abstract}
\noindent
A pseudo-spin model is proposed, as a means to describe some transport
properties (resistivity and Hall mobility) in $%
Bi_2Sr_2(Ca_zPr_{1-z})Cu_2O_{8+y}$. Our model is based in a double-well
potential where tunneling in a given site and interaction between different
lattice sites are allowed only through the excited states. Doping of the
pure system by the addition of $Pr$ increases the ratio between the
activation energy and the tunneling constant. The model Hamiltonian displays
some features which are present in the hydrogen-bonded ferroelectrics. Its
dynamics is treated in the random phase approximation and the characteristic
frequency (time) is used in a Drude formula in order to obtain some
transport properties of the system, namely the electric resistivity and the
Hall mobility. The quantities calculated in this work are compared with the
experimental data of B. Beschoten, S. Sadewasser, G. G\"{u}ntherodt and C.
Quitmann [Phys. Rev. Lett.77, 1837(1996)].\\[5mm]
\end{abstract}

Pacs number: 74.20.De, 74.25.Dw, 74.25.Fy, 74.72.Hs

\section{Introduction}

The controversy behind the mechanism which governs high temperature
superconductivity (HTS) in cuprates is a subject of current interest, as can
be verified in the recent literature [1,2,3].

Anderson [2] attributes the novel phenomenology present in cuprates
materials to a second kind of metallic state, namely, the Luttinger liquid.
Zhang [3] has proposed a SO(5) theory of cuprate superconductivity which
considers that the phenomenology of these materials might be fundamentally
due to a conflict between different kinds of order. Chianchi, Moretti and
Piazza [4] have presented a phenomenological model for the formation of
Cooper pairs in cuprate superconductors: in their description, coupling is
mediated by large anharmonic oscillations along $\widehat{c}$ direction of
the apical ions. On the other hand, Beschoten et al [5] have investigated
the charge transport in insulating and superconducting samples of $%
Bi_2Sr_2(Ca_zPr_{1-z})Cu_2O_{8+y}$, in terms of the Hall mobility. The main
motivation behind Beschoten et al work [5] was that the interplay of
localization and superconductivity is of fundamental interest in the physics
of electrons (holes) in strongly correlated and disordered systems such as
the high-temperature superconductors (HTSC).

The aim of this paper is to propose a phenomenological model to explain the
resistivity and the Hall mobility behaviors inferred from Beschoten et al
data [5], performed in $Bi_2Sr_2(Ca_zPr_{1-z})Cu_2O_{8+y}$. The strategy of
our work is first to propose a model Hamiltonian to describe the $%
Bi_2Sr_2(Ca_zPr_{1-z})Cu_2O_{8+y}$ and then to study its dynamics in the
random phase approximation (RPA). After doing that, we use the obtained
characteristic frequency (time) in the Drude formula, in order to obtain the
resistivity of the system. The Hall mobility will also be studied by using
the above results.

Gaona and Silva [6], inspired in the Gorter-Casimir (GC) two-fluid model of
superconductivity [7], as well in the formalism used to treat the phase
transitions in hydrogen-bonded ferroelectrics [8], have proposed an
effective transverse Ising model (TIM) to describe both metallic and HTS
superconductivity. We think that an extension of this previous model could
be useful to describe some basic features of Beschoten et al [5] experiment.

Let us first make some considerations about two well-established
phenomenological theories of the superconductivity. One of the main
characteristics of the superconducting state is the macroscopic phase
coherence of the wave function. In 1950, Ginzburg and Landau [9] introduced
their phenomenological theory of superconductivity, where the order
parameter must be identified with the macroscopic wave function $\Psi $.
They started from the idea that $\Psi $ represents some effective wave
function of the superconducting electrons. $\Psi $ can be normalized in such
a way that $\left| \Psi \right| ^2$ is equal to the concentration $n_s$ of
superconducting electrons. It seems that the pioneering theory to treat the
thermodynamic behavior of the superconducting state was proposed by Gorter
and Casimir (GC) [7]. The two-fluid model of Gorter and Casimir was proposed
to provide a basis for understanding the thermodynamic of the
superconducting state. In this phenomenological theory it is assumed that
the electrons in a superconductor are divided into two interpenetrating
gases: the superconducting electrons, which have zero entropy, and the
normal electrons, which have the usual properties of electrons in a normal
metal. The normal electrons give up the condensation energy $\beta $ when
they become superconducting. The authors proposed that below $T_c$ a
fraction $y$, where $y$ depends on the temperaure $T$ of the conduction
electrons occupy a set of  lower ''condensed'' energy states which are
associated with the superconducting properties, while the fraction $(1-y)$
remains uncondensed. The free energy function for the whole electronic
assembly is made up of two terms: one, $F_N$, is similar to the electronic
free energy of a normal metal and the other, $F_S$, corresponds to the
condensation energy associate with the condensed state , 
\begin{equation}
F_N=-\frac 12\gamma T^2\text{, }F_S=-\beta 
\end{equation}
where $\gamma $ is the normal electronic specific heat coefficient, and $%
\beta $ is a constant representing the condensation energy. Gorter and
Casimir assumed that the phases were not independent, and proposed the
following free energy function: 
\begin{equation}
F\left( T,y\right) =\left( 1-y\right) ^{1/2}F_N+yF_S\text{ .}
\end{equation}

Close to $T_{c\text{ }}$we may express (2) in the standard Ginzburg-Landau
form by expanding $F\left( T,y\right) $as a power series in $y$, assuming $%
y=\left| \Psi \right| ^2$, and retaining the first nonvanishing terms. The
model has acquired support by its success to interpret all properties of
superconductors. In particular, the temperature dependence of the
penetration depth is correctly predicted [10,11].

An inspection of the GC free energy relation (2) reveals that it contains
two contributions: the first one is related to the property of free
electrons in metals to transpose the Fermi barrier, being responsible for
the transport properties and consistent with the Pauli exclusion principle.
This term also contains a multiplicative factor which takes into account the
fact that, below $T_c$, the number of normal electrons in the
superconducting sample is a temperature dependent quantity. The second term
of the GC free energy is responsible for the pairing of wave functions of
the electron condensate. This term is also temperature dependent (indeed, it
is proportional to the number of the electrons in the condensate) and above $%
T_c$ it does not contribute to the GC free energy. So, the GC free energy
clearly displays the competition between these two terms.

On the other hand, a very successful model to treat various cooperative
phenomena is the Ising model. However, the Ising model suffers from a
deficiency, namely, the impossibility to exhibit a proper dynamics. Perhaps
the simplest extension of the Ising model with a proper dynamics is the
transverse Ising model (TIM). The TIM has been used to describe phase
transition in ferroelectrics, ferromagnets and cooperative Jahn Teller
systems [12]. This model Hamiltonian was first proposed by De Gennes [13]
based on arguments developed by Blinc [14] to represent the basic features
of hydrogen-bonded ferroelectrics of the KH$_2$PO$_4$ family. In these
systems the Ising term corresponds to the interaction between the protons at
different lattice sites and the transverse field accounts for the
possibility of protons occupying one of the two minima of a double potential
well in a given site.

\section{The Model Hamiltonian}

Taking into account the preceding considerations, we propose the following
effective Hamiltonian as a means to describe some transport properties of
the $Bi_2Sr_2(Ca_zPr_{1-z})Cu_2O_{8+y}$. We write: 
\begin{equation}
H=-\Delta \frac 12\mathrel{\mathop{\sum }\limits_{i}}C_i-\Omega %
\mathrel{\mathop{\sum }\limits_{i}}\frac{1-C_i}2X_i-\frac 12%
\mathrel{\mathop{\sum }\limits_{ij}}L_{ij}\frac{1-C_i}2Z_i\frac{1-C_j}2Z_j%
\text{ .}
\end{equation}

The main ideas that we have in mind are:

i-) The third term of (3) is an Ising-like term, where the operator $Z$ is
related to the wave function of the condensate electrons (holes), and the
coupling $L_{ij}$ favors the pairing of electrons (holes) at two different
lattice sites, contributing to the coherence of their wave functions below
the critical temperature $T_c$.

ii-) In the second term of (3) the operator $X$ is related to the wave
function of the normal electrons (holes), where $\Omega $ represents the
transverse field or tunneling constant. This term, which gives dynamics to
the model, could be thought as the kinetic contribution to the model
Hamiltonian. It could also represent a mimic for the motion of ''free''
electrons (holes) through the Fermi barrier, in close analogy with the
tunneling of protons through the double-well potential barrier in the KH$_2$%
PO$_4$-like (hydrogen-bonded ferroelectrics) case (see [8]). However, while
in KH$_2$PO$_4$ the proton tunneling occurs between the right and left sides
of the potential well, in the present model we consider the possibility of
the electron (hole) to tunnel between the two discrete values of the phase $%
\varphi $ of its wave function, namely between $\varphi =0$ and $\varphi
=\pi $. According to this point of view, the spontaneous symmetry breaking
will occurs when the system makes the choice between one of these two
particular phases of the wave function.

iii-) In the first term of (3), $\Delta $ represents the metal-insulator
activation energy, and the operator $C$ will assume the eigenvalues +1, when
the electron (hole) occupies the ground state (insulator), and -1, when it
occupies the excited state (metal or superconductor). We observe that the
electron (hole) could change the phase of its wave function only in the -1
eigenstate of $C$. The interaction of electrons (holes) between different
lattice sites, which leads to the cooperative effect of the
superconductivity, occurs only in this last case.

With respect to the operators which appear in the Hamiltonian (3), we would
like to make the following remarks: $X$ and $Z$ will be treated as linear
independent operators in the space of spin-1/2, described by the Pauli
matrices. The operator $C$, which has eigenvalues +1 and -1, could be
treated as a '' z component '' of a spin-1/2 operator in another space. In
this way, the Hamiltonian (3) is defined in the space product of the
operators $C$ and ( $X,Y,Z$ ).

Before finishing this section, it is worth to mention that a Hamiltonian of
the kind described by (3) was used by Ohtomi and Nakano [15] to explain the
role of the pressure in determining the order of the phase transition in
hydrogen-bonded ferroelectrics. They however treated (3) only in the static
case, namely in the mean field approximation (MFA).

\section{The Dynamics of the System}

The Dynamics of the system will be treated by applying the RPA to the
Heisemberg equations of motion. We have (with $\hbar =1$) 
\begin{equation}
\frac{d\left\langle P_i\right\rangle }{dt}=-i\left\langle \left[
P_i,H\right] \right\rangle _t\text{ ,}
\end{equation}
where the operators $P_i=X_i$, $Y_i$, $Z_{i\text{ }}$must satisfy the Pauli
commutation relations, 
\begin{equation}
\left[ X_i,Y_i\right] =i\delta _{ij}Z_i\text{, ... .}
\end{equation}
To solve (4) in the RPA, we write 
\begin{equation}
\left\langle P_i\right\rangle _t=\left\langle P_i\right\rangle +\delta
\left\langle P_i\right\rangle e^{i\omega t}\text{,}
\end{equation}
where in (6) we make the replacement of $\left\langle P_i\right\rangle _t$
by a constant part $\left\langle P_i\right\rangle $, which is just the mean
field approximation (MFA) expected value, plus a small time-dependent
deviation $\delta \left\langle P_i\right\rangle e^{i\omega t}$.

Using (5) and (6), keeping only terms which are linear in the deviation $%
\delta \left\langle P_i\right\rangle $ and by putting $\left\langle
Y\right\rangle =0$, we have the following set of linearized equations:

\[
i\omega \delta \left\langle X\right\rangle -L_0\left( \frac{1-\left\langle
C\right\rangle }2\right) \left\langle Z\right\rangle \delta \left\langle
Y\right\rangle =0 
\]
\begin{equation}
\text{ }i\omega \delta \left\langle Y\right\rangle -\left[ \Omega -L_0\left( 
\frac{1-\left\langle C\right\rangle }2\right) \left\langle X\right\rangle
\right] \delta \left\langle Z\right\rangle +L_0\left( \frac{1-\left\langle
C\right\rangle }2\right) \left\langle Z\right\rangle \delta \left\langle
X\right\rangle =0
\end{equation}

\[
i\omega \delta \left\langle Z\right\rangle +\Omega \delta \left\langle
Y\right\rangle =0 
\]

The solutions of (7) are: 
\begin{equation}
\omega _1=0
\end{equation}
and 
\begin{equation}
\omega _{2,3}^2=\Omega \left[ \Omega -L_0\left( \frac{1-\left\langle
C\right\rangle }2\right) \left\langle X\right\rangle \right] +\left[
L_0\left( \frac{1-\left\langle C\right\rangle }2\right) \left\langle
Z\right\rangle \right] ^2\text{,}
\end{equation}
which are constrained by the zero-order solution 
\begin{equation}
\left[ \Omega -L_0\left( \frac{1-\left\langle C\right\rangle }2\right)
\left\langle X\right\rangle \right] \left\langle Z\right\rangle =0\text{ .}
\end{equation}

In the following we are interested only in the characteristic frequency of
the electrons (hole) in the non-superconducting phase, i.e., in the phase
given by $\left\langle Z\right\rangle =0$ .

Putting $\left\langle Z\right\rangle =0$ in $\omega _{2,3}^2$ given by (9)
and dropping the subscripts for sake of simplicity we have: 
\begin{equation}
\omega ^2=\Omega \left[ \Omega -L_0\left( \frac{1-\left\langle
C\right\rangle }2\right) \left\langle X\right\rangle \right]
\end{equation}
which leads to: 
\begin{equation}
\omega _{\pm }=\pm \left\{ \Omega \left[ \Omega -L_0\left( \frac{%
1-\left\langle C\right\rangle }2\right) \left\langle X\right\rangle \right]
\right\} ^{1/2}\text{.}
\end{equation}

In (12) the values of $\left\langle C\right\rangle $ and $\left\langle
X\right\rangle $, in the MFA, are given by 
\begin{equation}
\left\langle X\right\rangle =\frac 12tgh\left( \frac \Omega {2k_\beta T}%
\right)
\end{equation}
and 
\begin{equation}
\left\langle C\right\rangle =tgh\left( \frac{\Delta -\Omega \left\langle
X\right\rangle }{2k_\beta T}\right) \text{ .}
\end{equation}

\section{Transport Properties}

\subsection{Evaluation of the Resistivity}

To evaluate the resistivity we will work with the Drude formula,

\begin{equation}
\rho =\frac m{pe^2\tau }
\end{equation}
where $m$ is the effective mass of the carriers, $p$ their concentration, $e$
the electronic charge and $\tau $ the scattering time.

Let us now make the link between the dynamics of our model and the Drude
formula. Since the carriers concentration must be an increasing function of
the $Ca$ concentration, $f\left( z\right) $, and must be a maximum when all
the electrons (holes) occupies the excited state $\left( \left\langle
C\right\rangle =-1\right) $ and a minimum for the ground state $\left(
\left\langle C\right\rangle =+1\right) $, we propose: 
\begin{equation}
p=n\text{ }f\left( z\right) \left( \frac{1-\left\langle c\right\rangle }2%
\right) ^2\text{ ,}
\end{equation}
where $n$ is the density of carriers which are able to form cooper pairs. We
also assume [16] that 
\begin{equation}
n=\frac 2{\xi _{ab}^2\xi _c}\text{ ,}
\end{equation}
where $\xi _{ab}$ and $\xi _c$ are respectively the in-plane and the
perpendicular coherence lengths. The number 2 corresponds to the assumption
that in the superconducting phase we will have only a cooper pair by ''unit
cell'' of volume $\xi _{ab}^2\xi _c$.

By putting 
\begin{equation}
\tau =\frac 1{2\left| \omega _{\pm }\right| }
\end{equation}
in (15) we get 
\begin{equation}
\rho =\left( \frac{2m}{e^2}\right) \frac{\left| \omega _{\pm }\right| }p%
\text{ , }
\end{equation}
and putting (12) and (16) into (19) we get: 
\begin{equation}
\rho =\left( \frac{8m}{e^2\hbar }\right) \frac{\left\{ \Omega \left[ \Omega
-L_0\left( \frac{1-\left\langle C\right\rangle }2\right) \left\langle
X\right\rangle \right] \right\} ^{1/2}}{nf\left( z\right) \left(
1-\left\langle c\right\rangle \right) ^2}\text{ .}
\end{equation}

One immediate consequence of relation (20) is that the onset of the
superconductivity is reached when the term in the square-root goes to zero.
This corresponds to the softening of the mode which frequency is given by
equation (12).

\subsection{Hall Mobility}

To make a connection between our model and the Hall mobility $\mu $, we will
consider the one-band model [17] where the Hall effect is inverse
proportional to the carrier concentration $p$:

\begin{equation}
R_H=\frac 1{pe}\text{ .}
\end{equation}

Using this assumption we will have for the mobility

\begin{equation}
\mu =\frac{R_H\left( T,z\right) }{\rho \left( T,z\right) }=\frac e{2m\left|
\omega _{\pm }\right| }\text{ ,}
\end{equation}
where we have used (19) to obtain the right side of (22).

Putting (12) into (22) we finally obtain 
\begin{equation}
\mu =\frac{\left| e\right| \hbar }{2m}\left\{ \Omega \left[ \Omega
-L_0\left( \frac{1-\left\langle C\right\rangle }2\right) \left\langle
X\right\rangle \right] \right\} ^{-1/2}\text{ .}
\end{equation}

\subsection{A Possible Mode of Vibration}

We also can evaluate the resistivity with an equivalent way of writing the
Drude formula

\begin{equation}
\rho =\frac{mv_F}{pe^2l}.
\end{equation}
where $v_{F\text{ }}$ is the Fermi velocity and $l$ the mean free path of
the electrons (holes).

As a means to determine the electron (hole) mean free path we start with the
classical formula 
\begin{equation}
l=\frac 1{n\pi \overline{r^2}}\text{ .}
\end{equation}

We observe that in (25) we have considered the number of scatter centers by
unit of volume to be equal to the density of Cooper pairs given by (17). We
can justify this by the argument that the electric conduction always occurs
in a regime of charge neutrality.

The electron (hole) mean free path, $l$, remains to be linked to the model's
dynamics. Alternatively, we can work with the collision cross-section $\pi 
\overline{r^2}$ taking into account equation (25). To do this let us write
the ''resonance'' condition: 
\begin{equation}
\frac 12K\overline{r^2}=2\hbar \left| \omega _{\pm }\right| \text{ .}
\end{equation}

In (26), we have considered the equality in energy between a classical
harmonic oscillator of elastic constant $K$ and the energy separation of a
two-level quantum system, which frequencies $\omega _{\pm \text{ }}$are
given by (12).

Initially we are going to determine the elastic constant $K$ through the
following arguments. It seems natural to consider that the Debye frequency $%
\omega _{D\text{ }}$ and the mass $M$ of the oxygen ion must play an
important role in the scattering of the charge carriers. So, let us write
the relation: 
\begin{equation}
K=\alpha M\omega _D^2\text{ .}
\end{equation}

We can suppose that in (27) $K$ is an effective elastic constant, where $%
\alpha $ , an adjustable perhaps small pure number, could be interpreted as
a mimic for a weak coupling regime of the BCS [18] model.

Inserting the informations contained in the equations (25) and (26) into
(24), we obtain 
\begin{equation}
\rho =\left( \frac{4\pi n\hbar mv_F}{e^2K}\right) \frac{\left| \omega _{\pm
}\right| }p\text{ ,}
\end{equation}
and putting (12) and (16) into (28) we get: 
\begin{equation}
\rho =\left( \frac{16\pi mv_F}{e^2K}\right) \frac{\left\{ \Omega \left[
\Omega -L_0\left( \frac{1-\left\langle C\right\rangle }2\right) \left\langle
X\right\rangle \right] \right\} ^{1/2}}{f\left( z\right) \left(
1-\left\langle C\right\rangle \right) ^2}\text{ .}
\end{equation}

Since (19) and (28) are two equivalent ways of writing the resistivity, we
obtain 
\begin{equation}
K=2\pi n\hbar v_F\text{ .}
\end{equation}

On the other hand an estimate of the Fermi velocity $v_{F\text{ }}$was given
in reference [19], in terms of the electron (hole) mass and the in-plane
coherence $\xi _{ab}$, namely, 
\begin{equation}
v_F=\frac 32\frac \hbar {m\xi _{ab}}\text{ .}
\end{equation}

Putting (17), (27) and (31) into (30), we get 
\begin{equation}
K=\alpha \left( M\omega _D^2\right) =\frac{6\pi \hbar ^2}{\xi _{ab}^3\xi _cm}%
\text{ .}
\end{equation}

The fact that the right side of (32) depends only on the in-plane and
perpendicular coherence lengths and on the effective mass $m$ of the
carriers leads to the conjecture that it is possible to find a mechanism
accounting for the elastic constant $K$. In the following we are going to
look for this possibility.

In a previous work by one of the present authors [19], an effective
potential was proposed to describe the interaction between electron pairs
(holes) relating the size of this super carrier to the energy of the bound
state. By considering the particular situation that the particle occupies
the minimum of this effective potential and that the motion of the pair is
confined in a plane [20], we found that this problem could be mapped into
that of the quantum mechanical behavior of a particle described by polar
coordinates in a plane. The solution of this quantum mechanical problem
leads to an eingevalue spectrum which is doubly degenerate, except for the
ground state. In the first excited state we have two eingevalues of the
momentum $q$, namely: 
\begin{equation}
q_{\pm }=\pm \frac \hbar {\xi _{ab}}\text{ .}
\end{equation}

We may consider that a pair of electrons (holes) circling in opposite
directions establishes a ring of charge and that an oxygen ion could
experiment the electric field of these pair of carriers. This leads to an
oscillatory motion of the oxygen ion perpendicular to the plane, and in the
case of small oscillations, this motion could be described by a harmonic
oscillator which frequency $\omega _{0\text{ }}$is given by: 
\begin{equation}
M\omega _0^2=\frac{e^2}{\pi \varepsilon _0\xi _{ab}^3}\text{ .}
\end{equation}

Now, if we make the identification of the left side of (34) with the elastic
constant $K$ given by (32), we obtain: 
\begin{equation}
\xi _c=\left( \frac{4\pi \varepsilon _0}{e^2}\right) \frac{3\pi }2\frac{%
\hbar ^2}m\text{ .}
\end{equation}

So, according to our model the perpendicular coherence length $\xi _c$ is
inversely proportional to the effective mass $m$ of the charge carriers.

\section{Comparison with the Experiments and Numerical Estimates}

Let us now compare the experimental results of Beschoten et al [5] with the
present model. We may consider that the role played by the addition of $Pr$
in the $Bi_2Sr_2(Ca_zPr_{1-z})Cu_2O_{8+y}$ is to promote certain kind of
dilution which is responsible by the decreasing of the critical temperature
of the material. We assume that the dilution affects both the values of the
''tunneling constant'' $\Omega $, as well as the activation energy $\Delta $%
. In order to understand this, let us look at figure 1. In this figure we
have a double-well potential where the minimum corresponds to one of the two
values of the discrete phases of the local wave function, $\varphi =0$ or $%
\varphi =\pi $. The parameter $\Delta $ corresponds to the separation in
energy between the ground and the excited states. In figure 2 we show that
the effect of the tunneling is to make the splitting of the excited state in
two sub-levels. The system could become superconductor as far as $\Omega
/2>\Delta $. Therefore the critical concentration of $Ca$ ions ($z_c$),
i.e., the concentration below which we have no more the superconducting
state, will occur at $\Omega =2\Delta $.

Looking at the resisivity formula given by (20), we can observe that its
temperature dependence is controlled by the parameters of the model. So for $%
\Omega >2\Delta $, the onset of the superconductivity is reached when: 
\begin{equation}
\left\langle X\right\rangle _{T=T_c}=\frac{2\Omega }{L_0\left(
1-\left\langle C\right\rangle _{T_c}\right) }\text{ ,}
\end{equation}
where $\left\langle X\right\rangle $ and $\left\langle C\right\rangle $ are
given by (13) and (14), respectively.

In the numerical estimates of the resistivity (equation (20)), we have three
adjustables parameters: the tunneling constant $\Omega $, the activation
energy $\Delta $ and the ratio between the coupling and the tunneling
constants $L_0/\Omega $. Table 1 shows the values of the parameters adjusted
to reproduce the experimental data findings of Beschoten et al [5] . In
figure 3 we plot in a semi-logarithmic scale the resistivity (equation (20))
of the $Bi_2Sr_2(Ca_zPr_{1-z})Cu_2O_{8+y}$ as a function of the temperature
for various values of the dilution $z$. We used, in evaluate $n$ given by
(17), $\xi _{ab}=15\AA $\ and $\xi _c=3\AA $ [21], and $m$ is the electron
rest mass.\\[5mm]

\[
\begin{tabular}{ccccccccccccccc}
\hline
& $z$ &  &  &  & $\Delta $ &  &  &  & $\Omega $ &  &  &  & $L_0/\Omega $ & 
\\ \hline
& 0.3 &  &  &  & 12.62 &  &  &  & 24.0 &  &  &  & 4.2 &  \\ 
& 0.4 &  &  &  & 9.68 &  &  &  & 19.0 &  &  &  & 4.0 &  \\ 
& 0.5 &  &  &  & 7.61 &  &  &  & 15.0 &  &  &  & 3.8 &  \\ 
& 0.6 &  &  &  & 5.35 &  &  &  & 11.0 &  &  &  & 3.2 &  \\ 
& 0.7 &  &  &  & 1.31 &  &  &  & 2.8 &  &  &  & 4.1 &  \\ 
& 0.9 &  &  &  & 0.9 &  &  &  & 2.1 &  &  &  & 5.0 &  \\ \hline
\end{tabular}
\]
\\[5mm]

Table 1: Adjusted parameters values for various $Ca$ concentrations ($z$) in

$Bi_2Sr_2(Ca_zPr_{1-z})Cu_2O_{8+y}$ ($\Delta $ and $\Omega $ are in units of 
$10^{-21}J$).\\[5mm]

A linear relationship between $z$ and $\Delta /\Omega $ can be infered from
the data of table 1. From this relation we can estimate a value for the
critical concentration, $z_c$, approximately $0.51$, corresponding to $%
\Delta /\Omega =1/2$.

In the numerical estimates of the resistivity for the $%
Bi_2Sr_2(Ca_zPr_{1-z})Cu_2O_{8+y}$, in the version of equation (29), we have
used $\alpha $ as an adjustable parameter. $M$ is the mass of the oxygen
ion, $\omega _{D\text{ }}$is the Debye frequency given by $k_\beta \theta
_D/\hbar $, where we used for the Debye temperature $\theta _D=300K$ [22].
We have adjusted $\alpha =0.005$, consistent with the value of $M\omega _0^2$
obtained in (34).

We also obtain for $\xi _c$, using equation (35), the value of approximately 
$2.5\AA $.

Beschoten et al [5] have investigated the inverse of the Hall mobility, $\mu
^{-1}$, in $Bi_2Sr_2(Ca_zPr_{1-z})Cu_2O_{8+y}$ cuprate. They found that the
mobility of the conduction holes evolves from the insulator $(z<z_c=0.52)$
to the high temperature superconductor ($z>z_c$) by using a one-band model
with a single scattering time in order to fit the experimental data. Their
main result is that, even for high Ca content, $z>z_c$ , where the system
becomes a superconductor, they still clearly observe a minimum in the
inverse mobility, although its resistivity is purely metallic (this
indicates that, even in these samples, the superconducting wave function is
formed by states which are spatially localized).

However in our results we can not observe those minima in the inverse
mobility (equation (23)) for $z>z_c$, although they are present for $z<z_c$.

In figure 4 we use the zeros of the inverse Hall mobility and the minimum of
it ( for $z<z_c$) as a mean to construct a diagram separating the various
regimes: insulating, superconducting and metallic. This diagram agree with
the experiments (figure 4 of reference [5]) unless for the fact that it does
not reproduce the coexistence of superconducting and localization
(insulating).

\section{Concluding remarks}

It seems that the model proposed in this paper can capture the main features
which are present in the transport properties of the $%
Bi_2Sr_2(Ca_zPr_{1-z})Cu_2O_{8+y}$. In particular, we have evaluated the
resistivity which agree with the experimental data of Beschoten et al [5].

From the evaluated inverse Hall mobility we extracted a diagram $T$
(temperature) versus $z$ (concentration of $Ca$ ions) showing the various
regimes of the system, namely metallic, localized (insulating) and
superconducting regions.

By comparing two alternative ways of writing the Drude formula of
resistivity, and by finding a possible mode of vibration (equation (34))
very relevant to the problem, we can led to a relation which links the
effective mass of the carrier (hole) to the perpendicular correlation length 
$\xi _c$ (equation (35)). \\

The present research has been supported partly by CNPq-Brazil.

\newpage 

\begin{figure}[h]
\vspace{0.8cm}
\centerline{\epsffile{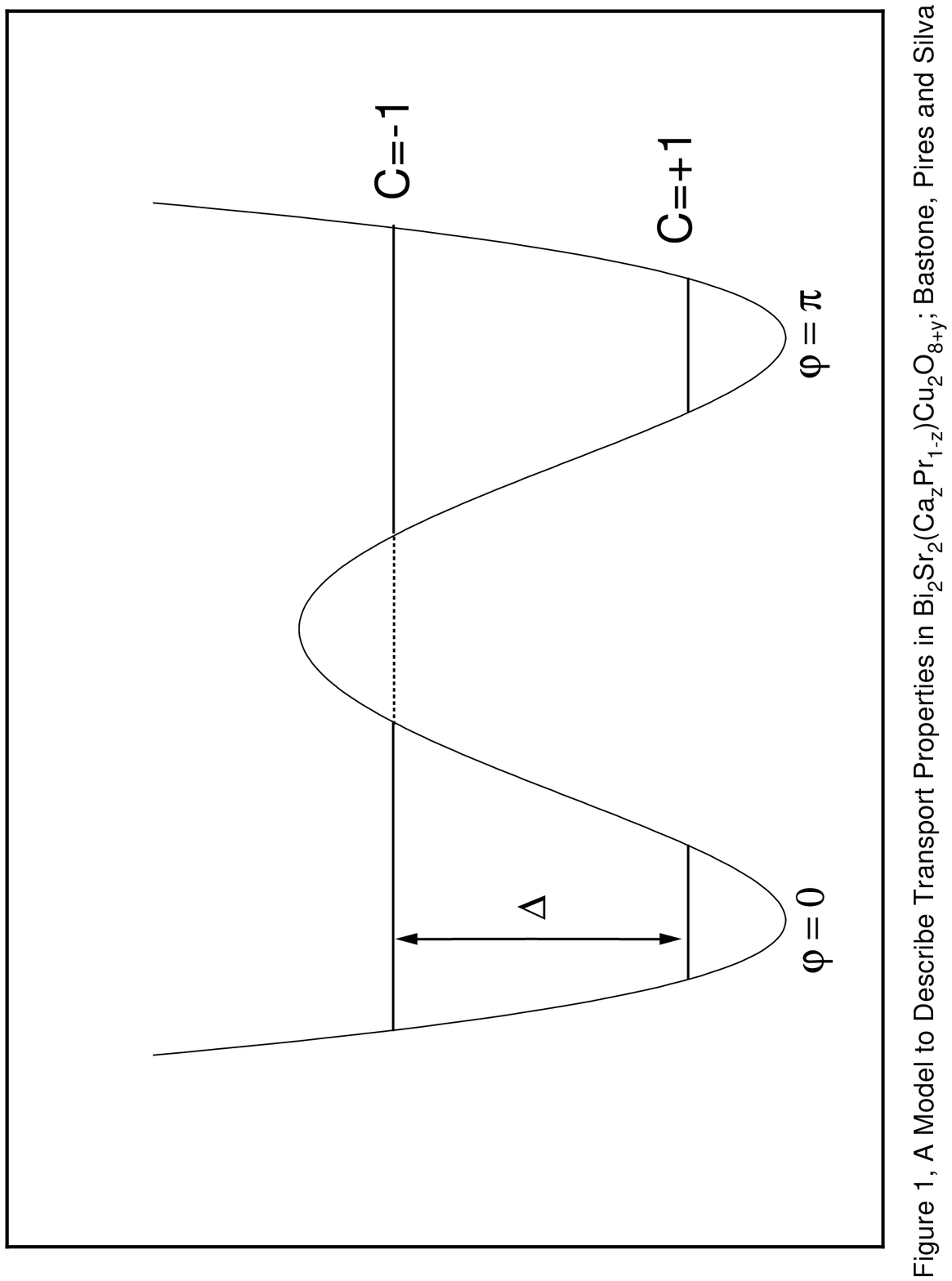} }
\vspace{0.8cm}
\protect\caption[]{
Figure 1: The double-well potential, showing the ground and first excited
states. At the minimal, the local phase of the wave function is $\varphi =0$
or $\varphi =\pi $.
}
\label{double-well}
\end{figure}

\begin{figure}[h]
\vspace{0.8cm}
\centerline{\epsffile{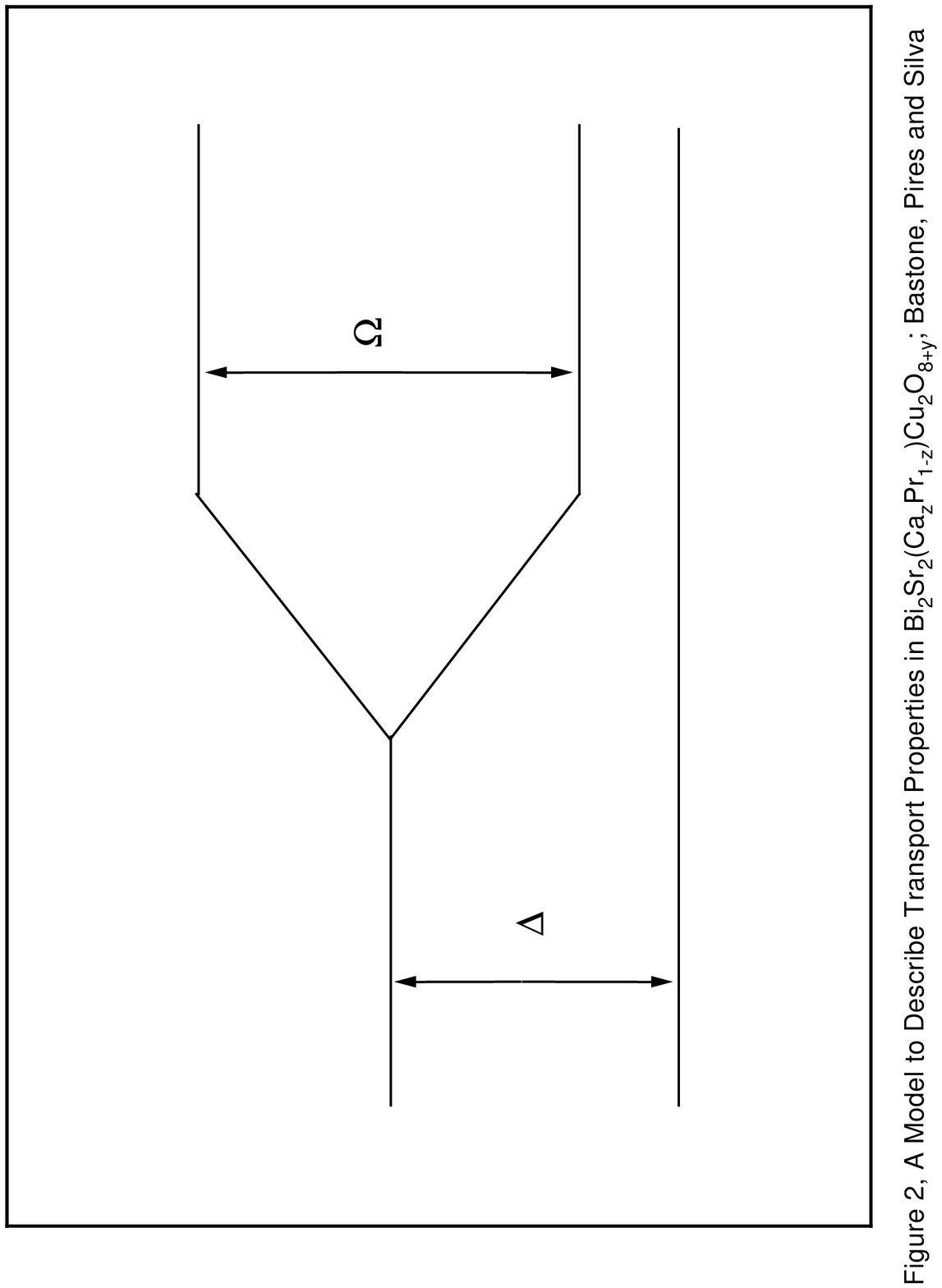} }
\vspace{0.8cm}
\protect\caption[]{
Figure 2: Splitting of excited state of the potential well, due to the
tunneling term $\Omega $. When $\Omega /2\geq \Delta $ the system is in the
superconductor regime (for $z>z_c$).
}
\label{splitting}
\end{figure}

\begin{figure}[h]
\vspace{0.8cm}
\centerline{\epsffile{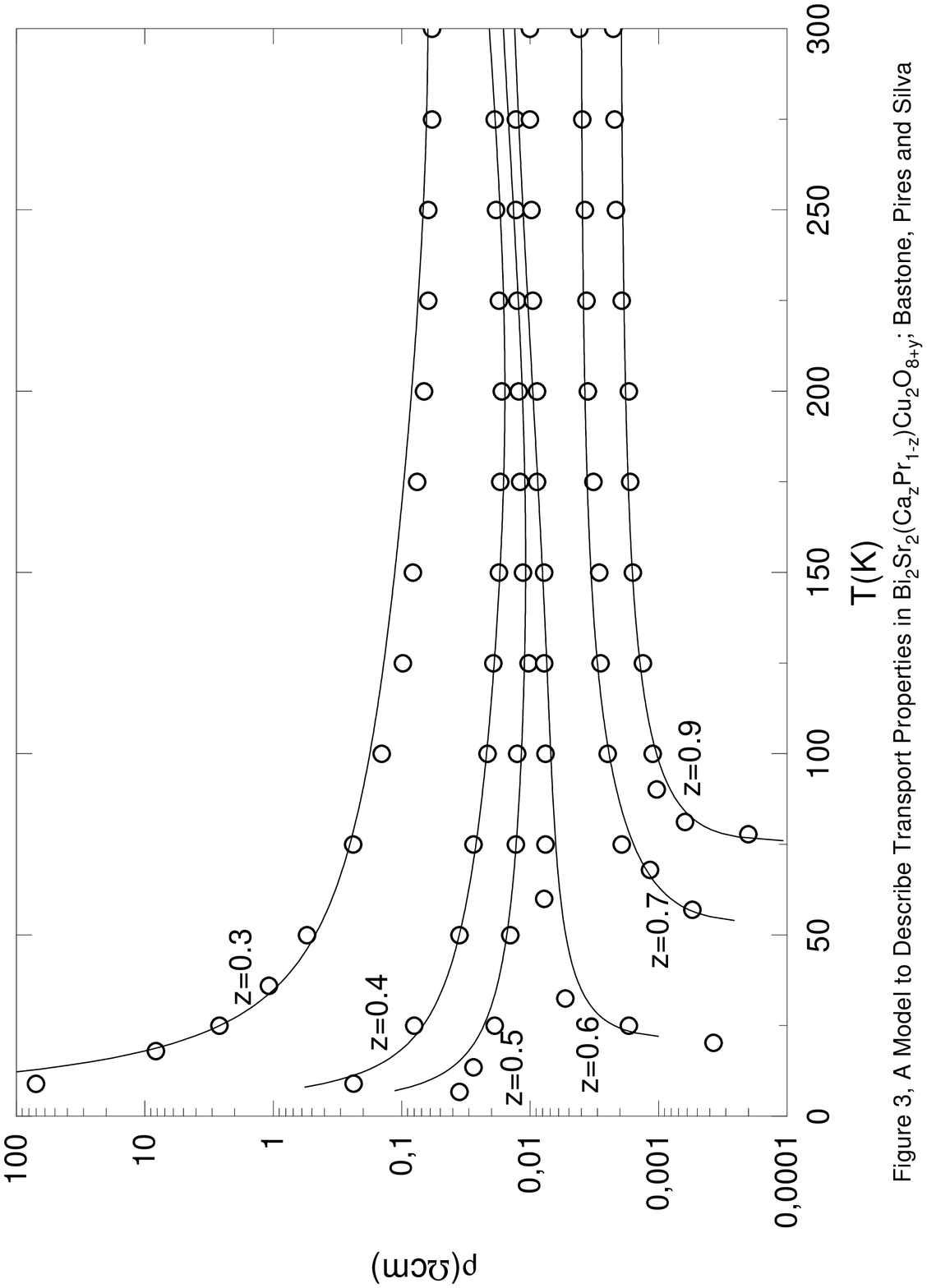} }
\vspace{0.8cm}
\protect\caption[]{
Figure 3: Logarithm of the resistivity as a function of the temperature, for
various values of of $Ca$ concentration ($z$). Theoretical results of this
work (solid lines) are compared with the experimental ones (circles) of
Beschoten et al [5].
}
\label{resistivity}
\end{figure}

\begin{figure}[h]
\vspace{0.8cm}
\centerline{\epsffile{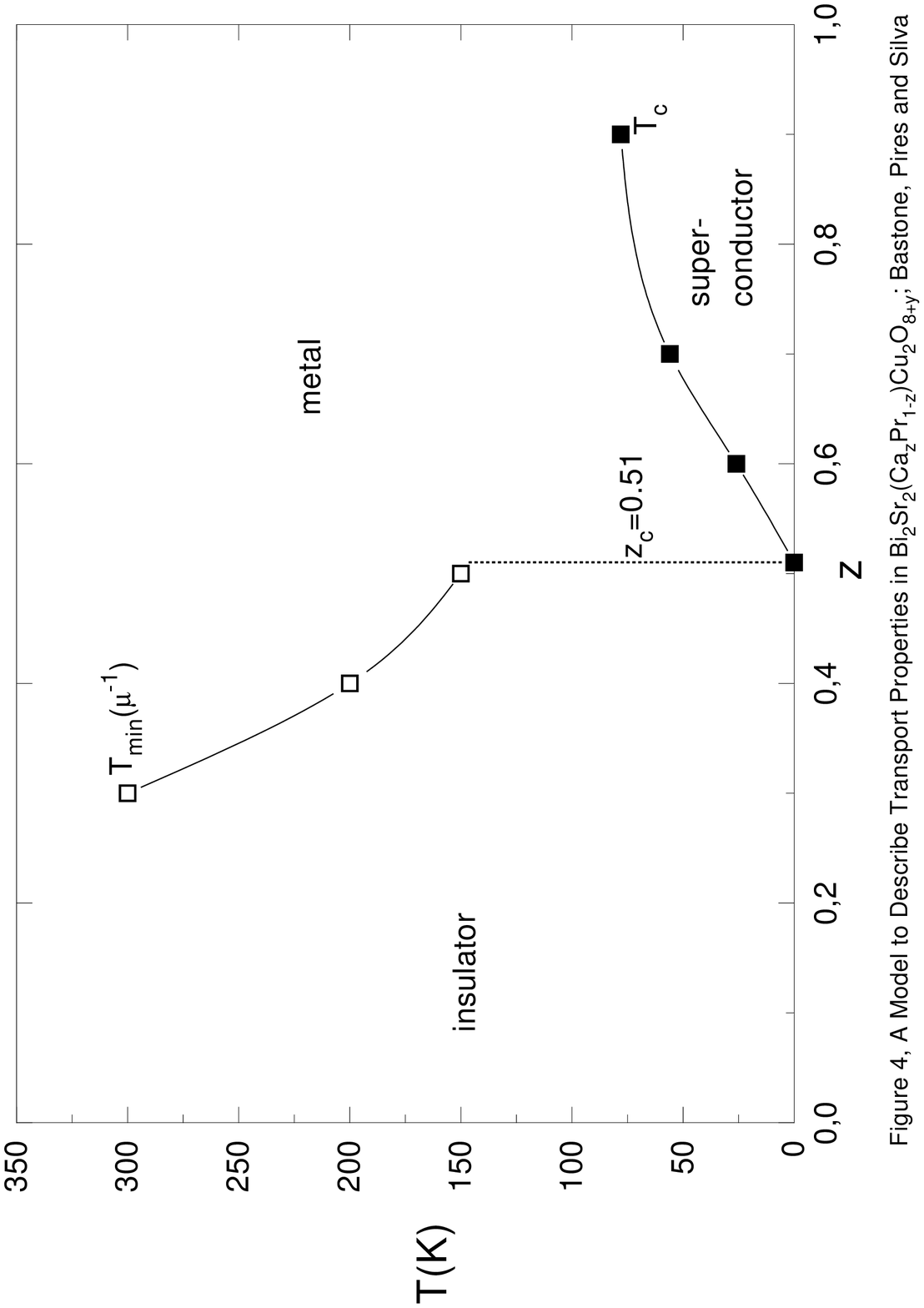} }
\vspace{0.8cm}
\protect\caption[]{
Figure 4: Inverse mobility minima corresponding temperature $T_{\min }$
(open squares) and superconducting transition temperature $T_c$ (filled
squares) for $Bi_2Sr_2(Ca_zPr_{1-z})Cu_2O_{8+y}$.
}
\label{mobility}
\end{figure}

\end{document}